\documentclass[twocolumn,prl,aps]{revtex4-1}
\usepackage{graphicx}
\usepackage{epstopdf}
\usepackage{amsmath}
\usepackage{amsfonts}

\usepackage{color}

\begin{document}
\title{From Compact to Open Clusters in Systems with Competing Interactions}

\author{Jakub P\c ekalski}
\affiliation{Institute of Physical Chemistry, Polish Academy of Sciences
Kasprzaka 44/52, 01-224 Warsaw, Poland}
\author{Andrew P. Santos and Athanassios Z. Panagiotopoulos}
\email{Author to whom correspondence should be addressed. Electronic mail: azp@princeton.edu}
\affiliation{Department of Chemical and Biological Engineering, Princeton University, Princeton, New Jersey 08544, USA}

\date{\today}
\begin{abstract}
Colloidal particles, amphiphiles, and functionalized nanoparticles are examples of systems that frequently exhibit short-range attractions coupled with long-range repulsions. In this work we observe significantly different dependences of the solution osmotic pressure versus composition upon the formation of clusters in different systems: a decrease in the pressure {\it vs.} density slope when clusters form in attraction dominating systems while clusters formed in repulsion dominating systems do not effect the pressure. This thermodynamic quantity turns out to be much more sensitive in picking out different clustering characteristics than the overall aggregation curve, cluster shapes or the lifetimes of clusters. Our results have significant implications in developing design principles for stable cluster self-assembly and detection in both laboratory settings and in computer simulations. 

\end{abstract}
%\maketitle
\maketitle

%
% Introduction
%
Colloids, proteins and functionalized nanoparticles can self-assemble into larger structures of finite size when the effective pair interactions are either highly anisotropic \cite{whitesides:02:0,wang:12:0, akcora:09:0} or are isotropic with competing attractive and repulsive forces at different distance ranges \cite{seul:95:0,rauh:17:0, stradner:04:0}. Amphiphiles, surfactants and block copolymers also self-assemble into aggregates commonly referred to as ``micelles'' because of the asymmetric nature of the molecules \cite{bates:99:0,matsen:96:0,santos:16:0}.  Structural and dynamic properties of systems which self-assemble into clusters due to competition between short-range attraction and long-range repulsion (SALR)  have been the subject of many recent studies \cite{mani:14:0,zhuang:16:0,toledano:09:0,candia:06:0,imperio:04:0,archer:07:0,zhang:09:0,chacko:15:0,sciortino:04:0,schwanzer:16:0,almarza:14:0,pekalski:14:0}.  Inhomogeneities arise due to the competition between the energy-driven preference to organize and the associated loss of the entropy in both amphiphilic and SALR systems \cite{seul:95:0,pekalski:14:1, ciach:13:0}.  Interestingly, it was shown that the two classes of self-assembling systems can be described by the same Landau-Brazovskii functional \cite{ciach:13:0}, and thus their mean-field phase diagrams are topologically equivalent at high temperatures. Such extraordinary similarity between SALR and amphiphilic systems provokes questions on whether the two systems stay similar when studied without resorting to mean-field approximations. 

Although clustering of SALR particles has been extensively investigated  \cite{sciortino:04:0,sciortino:05:0, toledano:09:0, klix:10:0, agarwal:09:0, tarzia:06:0,sweatman:14:0}, the osmotic pressure  versus density, $p(\rho)$, an important quantity frequently used to detect aggregation in surfactant systems \cite{amos:98:0,Floriano:99:0}, has not been previously reported. In this Letter, we report $p(\rho)$ for different attraction strengths and compare it with the characteristic behavior for micellizing systems. In such systems, the osmotic pressure follows the ideal-gas law when the concentration of amphiphiles is low and the system consists primarily of monomers. Micelles form above a certain concentration, known as the critical micelle concentration (cmc), at which point the slope of the osmotic pressure significantly decreases because the clusters are now the independent kinetic entities. The onset of micellization is also associated with the appearance of a separate peak in the mass-weighted distribution of the micelle sizes, the cluster size distribution (CSD), at the preferred aggregation number. The cmc is commonly defined as the concentration at which: the slope of the osmotic pressure changes, a preferred aggregation number appears and a maximum in the free oligomer concentration occurs \cite{santos:16:0}. Importantly, all these manifestations of structural change appear within a narrow concentration range. In this Letter, we show cases of SALR systems exhibiting traditional micellizing behavior and cases that differ in remarkable ways from this behavior.

In what follows, we first describe the model and methods. Secondly, we report the $p(\rho)$ and CSD in cases of attraction- and repulsion-dominated interactions. The osmotic pressure will hitherto be referred to as the ``pressure'', since they are equivalent in implicit-solvent systems. Finally, we attempt to understand the $p(\rho)$ and CSD  behavior by quantifying the static and dynamic properties of the self-assembled clusters. Our study of self-assembling SALR particles shows that, depending on the ratio of attraction and repulsion strength, clustering can be qualitatively very different than micellization.

%
% Model
%

We used the Lennard-Jones and  Yukawa potentials in order to model the SALR potential
\begin{equation}
V(r) = 4 \varepsilon \left[\left( \frac{\sigma}{r} \right) ^{2\alpha} - \left( \frac{\sigma}{r} \right) ^\alpha \right] + \frac{A}{r} e ^{-r/ \xi},
\end{equation}
where $\sigma =1$, $\alpha = 6$, $A = 0.5$ and $\xi = 2$; two values of $\varepsilon$ were considered: $\varepsilon = 1.6$ and $\varepsilon = 1.0$. The same form of the SALR potential, but with a much shorter range of attraction, was studied in Refs. \cite{sciortino:04:0,sciortino:05:0,mani:14:0,toledano:09:0}. In our case, after shifting the potential by its value at $r_{cut} = 8\sigma$ to zero, the attraction range is $2.042 \sigma$ (in the case of $\varepsilon = 1.6$) and $1.8096 \sigma$ (for $\varepsilon = 1.0$); thus, our model has similar ranges of attraction to those studied in Refs. \cite{zhuang:16:0,zhuang:16:1}. We chose the two values of $\varepsilon$ so that the second virial pressure coefficient, $B_2$, has a different sign over a significant range of temperatures below the Lennard-Jones critical temperature.

%
% Methods
%
We performed Metropolis Monte Carlo (MC) simulations \cite{metropolis:53:0} in the grand canonical ensemble using the Cassandra package \cite{Shah:11:0}, supplemented with a cluster center-of-mass displacement move algorithm \cite{Wu:92:0}. Two particles were considered clustered if the distance between them was less than the range of attraction. In order to reach equilibrium for each value of the chemical potential, $\mu$, we performed a sequence of simulations with decreasing temperature starting at $k_B T = 1.164 $  for $\varepsilon = 1.6$ and $k_B T = 0.831$ for $\varepsilon = 1.0$, where $k_B$ is the Boltzmann constant. In both cases, the temperature step was $0.0831$, and at each temperature $5 \times 10^6$ MC steps were performed, $10\%$ of which were the cluster moves, $50 \%$ were the single particle translations and $40\%$ were single particles insertions or deletions. Production runs consisted of $10^9$ MC steps after equilibration. The results were used in calculations of the pressure at a given temperature by histogram reweighting \cite{Ferrenberg:88:0,Ferrenberg:89:0,Floriano:99:0}.  To gain insight into the dynamical properties, we also performed molecular dynamics (MD) simulations in the canonical ensemble using the HOOMD-blue package \cite{hoomd:1,hoomd:2}. MD simulations were run using a timestep of 0.01 dimensionless time units with a Nos\'e-Hoover thermostat coupling constant of 10.  Production runs of $5 \times 10^6$ MD steps  were performed, starting with an equilibrated configuration from the MC simulations. A cubic box length of $17\sigma$ was used in all simulations.\\

\begin{figure}
\includegraphics[scale=1.0]{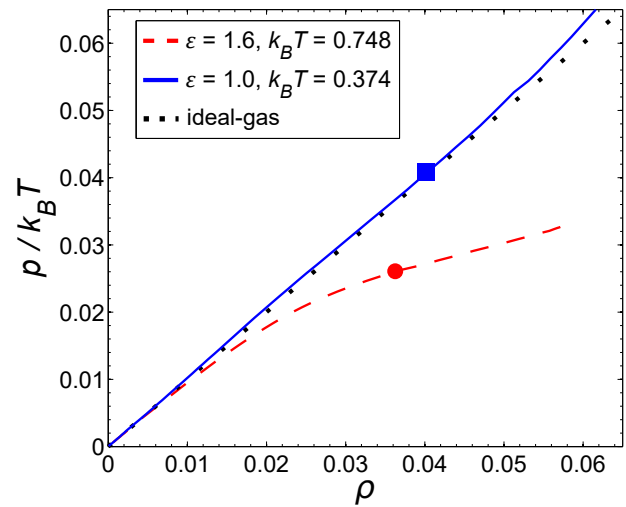}
\caption{Pressure vs. density for $\varepsilon = 1.6$ (dashed line), $\varepsilon = 1.0 $ (solid line) and the ideal-gas  equation of state (dotted line) . For $\varepsilon = 1.6$, attraction dominates and the slope changes above the critical cluster density; for $\varepsilon = 1.0$, repulsion dominates and the slope does not change. The red circle and the blue square correspond to the critical cluster density estimated from the cluster size distributions shown in Fig. \ref{csd}. }
\label{pres}
\end{figure}

\begin{figure}
\includegraphics[scale=1.0]{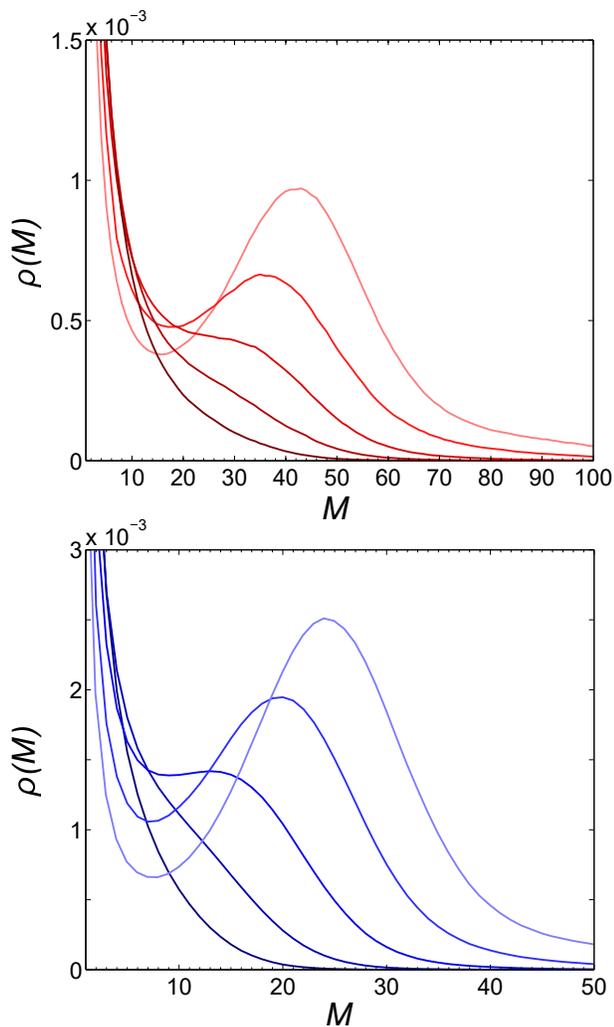}
\caption{Cluster size distribution (CSD) weighted by the average density, $\rho$, in the attraction-dominated case, $\varepsilon = 1.6$, at $k_BT = 0.748$ (upper panel) and the case of dominant repulsion, $\varepsilon = 1.0$, at $k_BT = 0.374$ (lower panel). In both panels, the lighter the color, the denser the system. The presented curves are from MC simulations with the following chemical potential values:  $\mu = 0.036, 0.0365, 0.037, 0.0375, 0.038$ (upper panel)  and $\mu = 0.023, 0.024, 0.025,0.026, 0.027$ (lower panel). The average densities change from $\rho = 0.0265$, to $\rho = 0.0532$ in the upper panel and from $\rho = 0.0242$ to $\rho = 0.0637$ in the lower panel. }
\label{csd}
\end{figure}

%
% Results
%
While the attraction between SALR particles leads to their aggregation, repulsion limits the size of the aggregates. In the attraction-dominated case, with $\varepsilon = 1.6$, the second virial coefficient $B_2 $ is negative for all temperatures $k_BT < 0.969$. Fig. \ref{pres} shows the pressure, $p$, as a function of the number density, $\rho$, for this system. The pressure obeys the ideal-gas law at low densities. Around a density of 0.02, the slope of $p(\rho)$  decreases significantly.  This decrease in the slope of the pressure coincides with the formation of clusters (see upper panel of Fig. \ref{csd}). The critical density at which clusters start to form, in analogy to the cmc, can be defined as the density at which a preferred size of the clusters occurs. Similar to micellization, the preferred aggregation number increases with increasing density, and the location of the local minimum is similar for different values of the chemical potential. At densities above the critical cluster density, $p(\rho)$ transitions to a straight line. The described clustering process strongly resembles micellization of amphiphilic molecules; thus, we conclude that clustering of SALR particles with an attraction dominating interaction is qualitatively very similar to micellization.\\

A repulsion-dominated potential does not prevent clustering in SALR systems.  In order to study such a case we set $\varepsilon = 1.0$, for which the second virial coefficient $B_2 > 0$ for $k_BT>0.266$. The lower panel of Fig. \ref{csd} shows that a separate peak in the CSD forms at $k_BT=0.374$.  The preferred aggregation numbers are smaller, compared to the attraction-dominated case, but nevertheless the CSD curves in both cases are qualitatively similar. According to the current understanding of surfactant micellization, the formation of a separate distribution of clusters, as seen at $k_BT = 0.374$, $\varepsilon =1.0$, should also correspond to a response in the pressure. However,  the slope of the pressure for this system (Fig. \ref{pres}) is almost the same as the ideal gas pressure; thus, the pressure does not respond to clustering. That is to say, SALR monomers and particles in clusters contribute equally to the pressure.

\begin{figure}[b]
\includegraphics[scale=1]{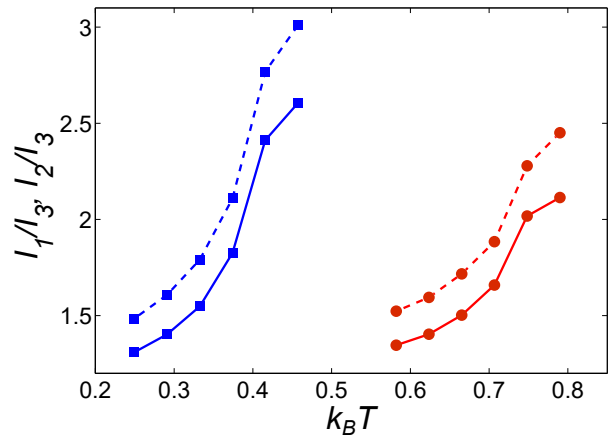}
\caption{Ratios between moments of inertia for different principle directions and for both attraction strengths. Red circles correspond to the $\varepsilon = 1.6$ case, while blue squares correspond to the $\varepsilon = 1.0$ case. Dashed lines connect results for $I_1 / I_3$, while solid lines connect results for $I_2/I_3$. The ratios were computed for preferred aggregation number-sized clusters at each temperature and density from MC simulations. }
 \label{mom}
\end{figure}

\begin{figure}
\centering
\includegraphics[width=0.4\textwidth]{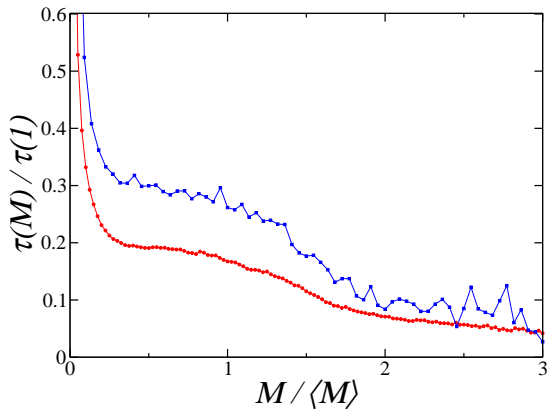}
\caption{ The mean life-times of clusters normalized by the monomer life-time, $\tau(M) / \tau(1)$, as a function of cluster sizes normalized by the preferred aggregation number, $M / \left\langle M \right\rangle$. The attraction-dominated case (red circles) exhibits qualitatively similar, but lower life-times than the repulsion-dominated case (blue squares). The attraction strengths and the temperature correspond to the $p(\rho)$ curves in Fig. \ref{pres}, $\rho = 0.053$. Preferred aggregation number for the attraction-dominated and repulsion-dominated cases were 40 and 22, respectively.}
\label{fig:lifetimes}
\end{figure}

Micelles formed by amphiphilic molecules affect the pressure because, in comparison to oligomers, i.e. aggregates smaller than the cluster size at which the CSD has a local minimum, micelles are bigger and thus diffuse more slowly \cite{viduna:98:0}.  In order to examine possible reasons why the self-assembled SALR clusters do not influence the pressure when repulsion is dominant, we studied  the structural and dynamic properties of the clusters for both values of $\varepsilon$. 
A structural description of the clusters can be made by considering their average size and shape. The size of the clusters is shown in Fig. \ref{csd}. Although particles in the repulsion-dominated system self-assemble into  smaller clusters, the small size of the aggregates does not explain the lack of the pressure response, because it has been shown that even small micelles can influence the pressure \cite{panag:02:1}. Fig. \ref{mom} shows that the average shape of the clusters changes with temperature, and that the higher the temperature, the less spherical the clusters. However,  the cluster shape does not show significant differences between the repulsion- and attraction-dominated cases.
The same is true when considering the cluster lifetimes. Following Ref. \cite{viduna:98:0}, we tracked the size of the aggregates during  each MD simulation. Whenever $m$ particles joined an aggregate of size $M$, an aggregate of size $M+m$  is born and an aggregate of size $M$ dies. Similarly, a detachment of $m$ particles from an aggregate of size $M$ is treated as the death of an aggregate of size $M$, and the birth of an aggregate of size $M-m$. We define the time from birth to death as the life-time of a cluster. With the above definitions, mean life-times of the aggregates, $\tau$, were calculated as a function of cluster size at densities well above the critical cluster density (Fig. \ref{fig:lifetimes}). The results show that the cases have qualitatively similar behavior.  Although, counter-intuitively, the attraction-dominated case affects the pressure, it has lower values of the cluster lifetimes, by this metric, than the repulsion-dominated case. We note that at lower temperatures, when the effect of the clustering on the pressure is strong, the average life-time of clusters is qualitatively similar to $\tau(M)$ found for block copolymers \cite{viduna:98:0}, specifically there is a maximum at the preferred cluster size and a minimum at the separation of oligomer and larger clusters.\\

Our finding shows that, for SALR systems, calculating the critical cluster density from thermodynamic properties, such as the pressure dependence on density, may not be sufficient for certain cluster-forming conditions, unlike traditional micellizing systems. Depending on the conditions, a transition in the cluster size distribution from being strictly monotonic to a distribution with a preferred aggregation number may not result in a decrease of the pressure slope. Future studies of this system will focus on the oligomer and cluster interactions and dynamics to help explain the unusual behavior detailed here.

Recent studies of SALR particles assume that the process of clustering is similar to micellization \cite{zhuang:16:0,zhuang:16:1}, and thus in order to find the critical cluster density one can use the methodology developed to determine the cmc for surfactant systems. Our results, however, show that one cannot identifying the critical cluster density based only on the pressure can miss certain clustering morphologies, such as the one identified here. On the other hand, critical cluster density calculations based only on the cluster size distributions do not provide comprehensive information about the system thermodynamics.

Experimental studies of SALR systems have yet to obtain periodic structures  \cite{stradner:04:0,jordan:14:0,campbell:05:0,klix:10:0,zhang:12:0,klix:13:0}. Our results suggest that  system design based solely on the desired structural characteristics of the aggregates can lead to the formation of entities which, regardless of their structure, may not be able to form a microphase, because of their dynamic instability.\\

J. P. acknowledges the financial support by the National Science Center: under Contract Decision No. DEC-2013/09/N/ST3/02551 and the funding of PhD scholarships on the basis of the Decision No. DEC-2014/12/T/ST3/00647.  A. Z. P. and A. P. S. acknowledge financial support from the Department of Energy, Office of Basic Energy Sciences, under Award No. DE-SC0002128. In addition, A. P. S. acknowledges funding from the National Science Foundation through a Graduate Research Fellowship.

\end{document}